\documentclass[final,5p,times,twocolumn]{elsarticle}
\usepackage{amssymb}
\usepackage{amsmath}
\usepackage{siunitx}
\usepackage{tabularx}
\usepackage{ifthen} % Allows custom conditions
\begin{document}

\begin{frontmatter}



\title{Redesign of the AD820 Single-Channel Circuit for the Development of the aRD820 Low-Noise Rail-to-Rail Operational Amplifier}


\author[1]{Dmitry Kostrichkin} %% Author name
\author[1]{Sergey Rudenko} %% Author name
\author[1]{Mihails Lapkis} %% Author name
\author[2]{Aigars Atvars \corref{cor1}} %% Author name
 \ead{Aigars.Atvars@lu.lv}

%% Corresponding author footnote
\cortext[cor1]{Corresponding author}
\affiliation[1]{organization={RD Alfa Microelectronics, Ltd.},%Department and Organization
            addressline={Latgales street 240}, 
            city={Riga},
            postcode={LV-1063}, 
            %state={},
            country={Latvia}}
\affiliation[2]{organization={Faculty of Science and Technology, University of Latvia},%Department and Organization
            addressline={Raina boulevard 19}, 
            city={Riga},
            postcode={LV-1586}, 
            %state={},
            country={Latvia}}

%% Abstract
\begin{abstract}
The task of this work was to design and later produce a low-power (single supply 5 - 30 V, dual supply $\pm$ 2.5 V and  $\pm$ 15 V) rail-to-rail operational amplifier aRD820 with low voltage noise ($<$4 $\mu$V, p-p. 0.1 to 10 Hz), ultralow input bias current ($<$ 15 pA), and low offset voltage ($<$ 500 $\mu$V) characteristics. Similar characteristics are presented by Analog Devices chip AD820. Thus, the task of the design team was to adapt the prototype circuity of AD820 to our technological capabilities, modify the circuit, if necessary, to eliminate any deficiency of the prototype. The input stage module got source followers at the input of the operational amplifier. Second stage module was modified to be more symmetric. The output stage module obtained additional resistors and capacitors to achieve a frequency compensation. One FET transistor in the current reference module was substituted by other elements. Simplified electric schemes of these modules of AR820 and aRD820 are presented. The performance of modified electric schemes of modules was tested in Simulink software. Simulations of the full electric scheme for aRD820 were made and showed that it demonstrates similar characteristics as AD820 data tables. Later production of the aRD820 chip and measurements demonstrated that the planned characteristics of the operational amplifier were met.
\end{abstract}

\begin{graphicalabstract}
\centering
\includegraphics[width=0.9\textwidth]{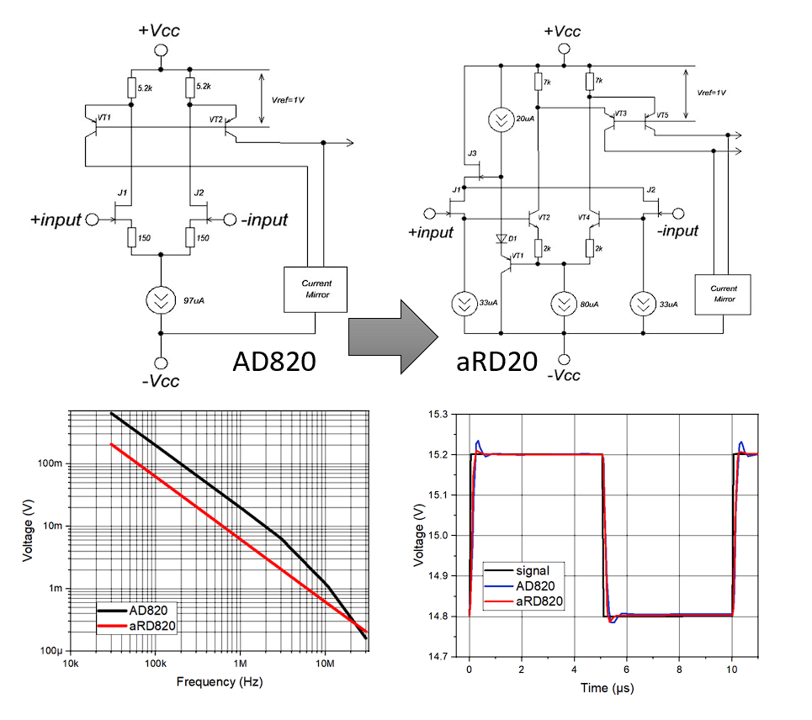}
\end{graphicalabstract}

\begin{highlights}
\item The task of this work was to design and later produce a low-power (single supply 5 - 30 V, dual supply $\pm$2.5 V and  $\pm$15 V) rail-to-rail operational amplifier aRD820 with low voltage noise ($<$ 4 $\mu$V, p-p. 0.1 to 10 Hz), ultralow input bias current ($<$ 15 pA), and low offset voltage ($<$ 500 $\mu$V) characteristics. Similar characteristics are presented by Analog Devices chip AD820. Thus, the task of the design team was to adapt the prototype circuity of AD820 to our technological capabilities, modify the circuit, if necessary, to eliminate any deficiency of the prototype
\item Direct replication of the electric scheme of AD820 and production of chips by our available production facility showed several drawbacks. For example, it was found that our N-channel FET transistors are inferior to the similar transistors used in the input stages and reference source circuits of the AD820. Specifically, amplifier stages using these transistors have poor noise characteristics. Additionally, the output voltage-current characteristic of our FET transistors has a broader "Ohmic region" compared to the FET transistors of the AD820. 
\item We are proposing an electric scheme for operation amplifier aRD820 that reach the planned characteristics and is obtained by modifying several modules of AD820 chip. The input stage module got source followers at the input of the operational amplifier. Second
stage module was modified to be more symmetric. The output stage module obtained additional resistors and capacitors to achieve a
frequency compensation. One FET transistor in the current reference module was substituted by other elements. For the production phase, we suggest using ion-implanted resistors instead of thin-film resistors. 
\item The performance of modules and full electric schemes of AD820 and aRD820 were
tested and compared in Microsim software. Simulations showed that aRD820 can reach expected parameters. Later production of the aRD820 chip and measurements demonstrated that the planned characteristics of
the operational amplifier are met.
\end{highlights}

%% Keywords

\end{frontmatter}



\section{Introduction}
\label{sec1}
%% Labels are used to cross-reference an item using \ref command.
An operational amplifier (op-amp) is an essential component in electronics that amplifies voltage \cite{Horowitz2015, Huijsing2017, Franco2015}. Op-amps are widely used in various applications like signal processing, filtering, and mathematical operations \cite{Jung2005, Yuce2024} because they can increase the strength of weak electrical signals and perform complex tasks with simple configurations.  While traditional operational amplifiers are designed to amplify voltage, operational transconductance amplifiers (OTAs) provide a current output, making them useful in several medical applications \cite{Manturshettar2019, Almalah2022}. CMOS operational amplifiers have gained widespread adoption in recent years due to their low power consumption and integration capabilities \cite{Sharma2024, Hussein2024}, yet amplifiers based on complementary bipolar technology \cite{Malhi1981, Bowers1999}, like the AD820, continue to excel in precision applications. Each technology serves distinct purposes: CMOS is ideal for modern, low-power digital systems, while bipolar amplifiers remain the gold standard for low-noise, high-precision analog applications. 

A rail-to-rail amplifier is a type of operational amplifier designed to utilize the full range of the power supply voltage, from the lowest supply rail (ground) to the highest supply rail (positive voltage). This means it can produce output voltages that go very close to the minimum and maximum supply voltages, enhancing its versatility in low-voltage applications. The development of low-voltage operational amplifiers capable of achieving rail-to-rail input and output operation has been a significant advancement in precision signal processing. One notable design achieves this functionality by incorporating a nested-loop frequency-compensation technique, allowing stable performance even at supply voltages as low as 1.5 V \cite{Huijsing1985}. Operational amplifiers are used also for high supply voltages, e.g. 30 V \cite{Wang2015}. For applications requiring rail-to-rail input and output, new bulk-driven amplifier designs offer significant performance improvements. The use of composite transistor arrays has been shown to provide superior voltage gain without sacrificing power efficiency \cite{Rodovalho2023}. A recent design using a folded cascode architecture and Class AB output stage achieved rail-to-rail performance with low power consumption and high gain \cite{Guang2012}. Another work proposes a rail-to-rail auto-zero operational amplifier using a time-interleaved charge pump circuit to ensure low power consumption and low offset voltage \cite{Zhang2024}.

Operational amplifiers could be single stage, two stage, three stage and multistage. The comparison of single and two stage CMOS op-amps showed that single stage op-amp is more stable and operate longer duration of time,  while two stage op-amp produces the larger output with lower noise \cite{Menberu2023}. 

The design of operational amplifiers for space applications requires consideration of radiation-induced effects, such as total ionizing dose (TID). Techniques such as hardened-by-design (RHBD) have been successfully implemented to mitigate these effects, as demonstrated in the development of a rail-to-rail operational amplifier resilient to radiation doses up to 500 krad(Si) \cite{Agostinho2016}.

Analog Devices developed a technology for vertical npn and pnp transistors known as complementary bipolar (CB) technology. This innovation significantly enhanced various chip solutions, including operational amplifiers. Leveraging this technology, Analog Devices introduced several operational amplifiers – single-channel AD820 (1993), two-channel AD822 (1994), and four-channel D824 (1995). These amplifiers exhibited exceptional characteristics regarding input current, noise levels, dynamic performance, and low power consumption. While they did not achieve the highest possible specifications in any single parameter, their balanced performance made them highly useful in many applications, particularly those requiring low input currents. 

The AD820 is a precision, low-power operational amplifier developed by Analog Devices, tailored for single-supply applications that demand high accuracy and efficiency. It operates over a wide supply voltage range (single supply 5 - 30 V, dual supply $\pm$ 2.5 V and  $\pm$ 15 V), making it suitable for both low-voltage portable devices and higher-voltage industrial systems. Key features include an ultralow input bias current of less than 15 pA and a low input offset voltage below 500 $\mu$V, which minimize errors when interfacing with high-impedance sources and enhance overall measurement precision. The amplifier also exhibits low voltage noise, specified at less than 4 $\mu$V peak-to-peak within the 0.1 Hz to 10 Hz frequency band, crucial for applications involving low-frequency, low-level signal amplification such as precision instrumentation and sensor signal conditioning. Moreover, the AD820 offers a rail-to-rail output swing, allowing the output voltage to approach the supply rails closely and thereby maximizing the dynamic range, especially in low-supply-voltage scenarios. Despite its low power consumption, with a typical quiescent current of only 800 $\mu$A, the amplifier maintains a moderate gain-bandwidth product suitable for a variety of analog signal processing tasks. Its design effectively minimizes both flicker (1/f) noise and thermal noise contributions, ensuring signal integrity across varying temperatures and supply voltages. These attributes make the AD820 an excellent choice for precision analog applications where low noise, ultralow input bias current, and low offset voltage are essential, including medical instrumentation, active filtering, and high-precision data acquisition systems.

The task of this work was to design and later produce a  low-power (single supply 5 - 30V, dual supply $\pm$ 2.5V and  $\pm$ 15 V) rail-to-rail operational amplifier with low voltage noise ($<$4 $\mu$V, p-p. 0.1 to10 Hz), ultralow input bias current ($<$ 15pA), and low offset voltage ($<$ 500 $\mu$V) characteristics. Similar characteristics are presented by Analog Devices chip AD820. Thus, the task of the design team of RD Alfa Microelectronics was to adapt the prototype circuity of AD820 to our technological capabilities, modify the circuit, if necessary, to eliminate any deficiency of the prototype, and utilize previous results obtained from designing the low-voltage four-channel amplifier microchip aRD824 \cite{Kostrichkin2022a, Kostrichkin2022b}. 

The list of planned parameters of aRD820  in comparison to datasheet values of AD820 is given in Table \ref{Table1}. It shows that aRD820 is expected to achieve very similar characteristics to AD820 and even proceed in some parameters.
%\clearpage
\begin{table*}[t]
    \centering
        \caption{Planned specification of an operational amplifier aRD820 chip and specification of Analog Devices AD820 chip. T$_{NORM}$ is +22 ± 3 $^o$C for aRD820 and +25 $^o$C for AD820. T$_{MIN}$ to T$_{MAX}$  is a temperature range (-60 $\ldots$ +125 $^o$C) for aRD820 and range (–40 $\ldots$ +85 $^o$C) for  AD820. Output saturation Voltage (High) is defined as the difference between the highest possible output voltage and the positive supply voltage. Output Saturation Voltage (Low) is defined as the difference between the lowest possible output voltage and the negative supply rail.}
    \label{Table1}
        
    %\scriptsize % to be commented in the final two-column version of the document
    \begin{tabularx}{\textwidth}{cXlcccccc}
       \hline   No.& Parameter & Temperature  &   \multicolumn{2}{c|}{aRD820 (planned)} &    \multicolumn{3}{c}{AD820 (AD820A) \cite{AD820_datasheet}} &  Units\\ \cline{4-8} 
        &  & & Min & Max & Min &  Typical & Max &  \\\hline 
        1 &  Offset Voltage&  T$_{NORM}$ & -0.5 & 0.5 &  &  0.1&0.8  &mV \\ 
           &  & T$_{MIN}$ to T$_{MAX}$& -1.5 & 1.5 &  & 0.5 & 1.2 & \\\hline 
         2& Input Bias Current &  T$_{NORM}$& -15 & 15 &  & 2 & 25 & pA\\
         &  & T$_{MIN}$ to T$_{MAX}$ & -4000 & 4000 &  & 500 & 5000 & \\\hline
            3& Input Offset Current &  T$_{NORM}$& -10 & 10 &  & 2 & 20 & pA\\
         &  & T$_{MIN}$ to T$_{MAX}$ & -500 & 500 &  & 500 & & \\\hline
        4 &  Large Signal Voltage Gain &  &  &  &  &  &  & V/mV\\
         & R$_L$ = 1 k$\Omega$ & T$_{NORM}$ & 10 &  &  &  &  & \\ \cline{3-8} 
         & R$_L$ = 2 k$\Omega$  &  T$_{NORM}$&  &  & 15 & 30 &  & \\
          &  &  T$_{MIN}$ to T$_{MAX}$&  &  &  10 &  &  & \\ \cline{3-8} 
          & R$_L$ = 10 k$\Omega$  &  T$_{NORM}$&  50&  & 80 & 150 &  & \\
          &  &  T$_{MIN}$ to T$_{MAX}$&  &  &  80 &  &  & \\ \cline{3-8} 
          & R$_L$ = 100 k$\Omega$  &  T$_{NORM}$& 250 &  & 400 &1000  &  & \\
          &  &  T$_{MIN}$ to T$_{MAX}$& 180 &  &  400 &  & & \\ \hline
          5 & Output Saturation Voltage (High)  &  &  &  &  &  &  & mV\\
        & I$_{SOURCE}$ = 20 $\mu$A  & T$_{NORM}$ &  & 14 &  &  10&14  &\\
        &  & T$_{MIN}$ to T$_{MAX}$ &  & 20 &  &  & 20 & \\ \cline{3-8} 
         & I$_{SOURCE}$ = 2.5 mA  & T$_{NORM}$ &  & 110 &  &  80&110  & \\
        &  & T$_{MIN}$ to T$_{MAX}$ &  & 160 &  &  & 160 & \\ \cline{3-8} 
        & I$_{SOURCE}$ = 15 mA  & T$_{NORM}$ &  & 1500 &  &  800&1500  & \\
         &  & T$_{MIN}$ to T$_{MAX}$ &  & 1500 &  &  & 1900 & \\ \hline
          6 & Output Saturation Voltage (Low)  &  &  &  &  &  &  & mV\\
        & I$_{SINK}$ = 20 $\mu$A  & T$_{NORM}$ &  & 7 &  &  5&7  & \\
        &  & T$_{MIN}$ to T$_{MAX}$ &  & 10 &  &  & 10 & \\ \cline{3-8} 
         & I$_{SINK}$ = 2.5 mA  & T$_{NORM}$ &  & 55 &  &  40&55  & \\
        &  & T$_{MIN}$ to T$_{MAX}$ &  & 80 &  &  & 80 & \\ \cline{3-8} 
         & I$_{SINK}$ = 15 mA  & T$_{NORM}$ &  & 500 &  &  300&500  &\\
        &  & T$_{MIN}$ to T$_{MAX}$ &  & 500 &  &  & 1000 & \\ \hline
           7  & Voltage noise, 0.1 Hz to 10 Hz & T$_{NORM}$ &  & 4 &  & 2 &  & $\mu$V, p-p\\ \hline
    \end{tabularx}
\end{table*}

\section{Inspection of AD820, replication, and drawback analyzes}
%% Use \subsection commands to start a subsection.
We inspected AD820 chip and derived its electric scheme. Then we replicated this electric scheme on a chip produced in our production facilities. Tests on several such chips showed drawbacks that did not allow to reach the performance of AD820. We made analyses on the drawbacks and got several conclusions. 

\begin{figure}
\centering
\includegraphics[width=0.7\columnwidth]{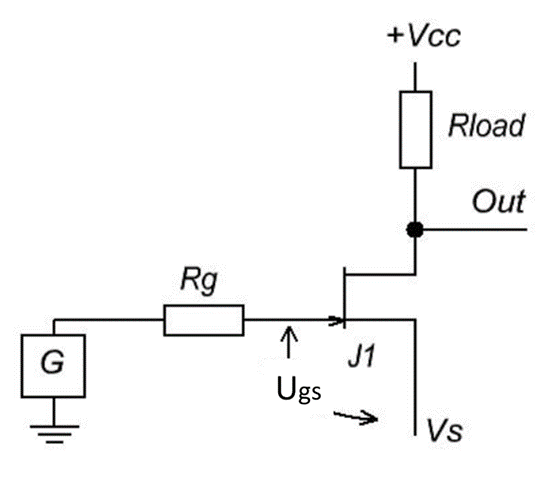}
\caption{Electric scheme to test FET transistors. +Vcc – supply voltage, Rg – generator resistor,  Rload – load resistor, J1 - FET transistor, G - signal generator, Vs – voltage on source, Ugs – voltage gate source.}
\label{Electric1}
\end{figure}

First, our N-channel FET transistors are inferior to the similar transistors used in the input stages and reference source circuits of the AD820. Specifically, amplifier stages using these transistors have poor noise characteristics. The noise voltage swing Vinp-p, referenced to the input, in the range of 0.1 Hz to 10Hz in the circuit shown in Figure \ref{Electric1}. is 5 $\mu$V$\ldots$10 $\mu$V, which is more than five times the corresponding value obtained in a similar circuit using an npn transistor. 

\begin{figure}
\centering
\includegraphics[width=\columnwidth]{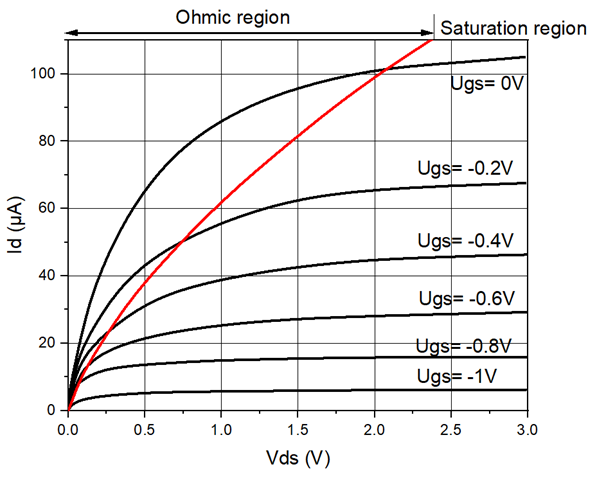}
\caption{Exit current-voltage characteristics FET transistor. Vds – voltage drain source, Id – drain current, Ugs – voltage gate source, Red line – separation between ohmic and saturation regions for various Ugs.}
\label{Ohmicregion}
\end{figure}

Second, the output voltage-current characteristic of our FET transistors has a broader "Ohmic region" compared to the FET transistors of the AD820. The initial section of the voltage-current characteristic of our FET transistor is shown in Figure \ref{Ohmicregion}. The "Ohmic region" is characterized by significantly lower transconductance (gm) values compared to the saturation region: $g_m=dI_d/dV_{gs}$, where $dI_d$ – increment of a drain current $I_d$, and $dV_{gs}$ – increment of a gate-source voltage $V_{gs}$.   If the FET transistor enters the "Ohmic region" (i.e., the source voltage is close to the drain voltage), the circuit's gain significantly decreases. Since the input stage of the AD820 is similar to the circuit in Figure \ref{Electric1}, its gain undergoes the same changes. When the input voltage ($V_{in}$) increases to +$V_{cc}$ -2V$\ldots$ + $V_{cc}$ -1V, the op-amp gain can drop to several hundred, especially in amplifiers with a higher gate cutoff voltage (in absolute value) $V_{gs0}$ (the gate source voltage at which $I_d$ becomes 0). Of two operational amplifiers with gate cut-off voltages of -2V and -1V, the gain reduction for input signals close to +$V_{cc}$ will be more significant in the amplifier with the gate cut-off voltage of -2V, since at equal $V_{in}$ values, the $V_{ds}$ value of these transistors will be lower by approximately 1 V (i.e., it will enter the "Ohmic region" earlier). 

\begin{figure}
\centering
\includegraphics[width=0.4\columnwidth]{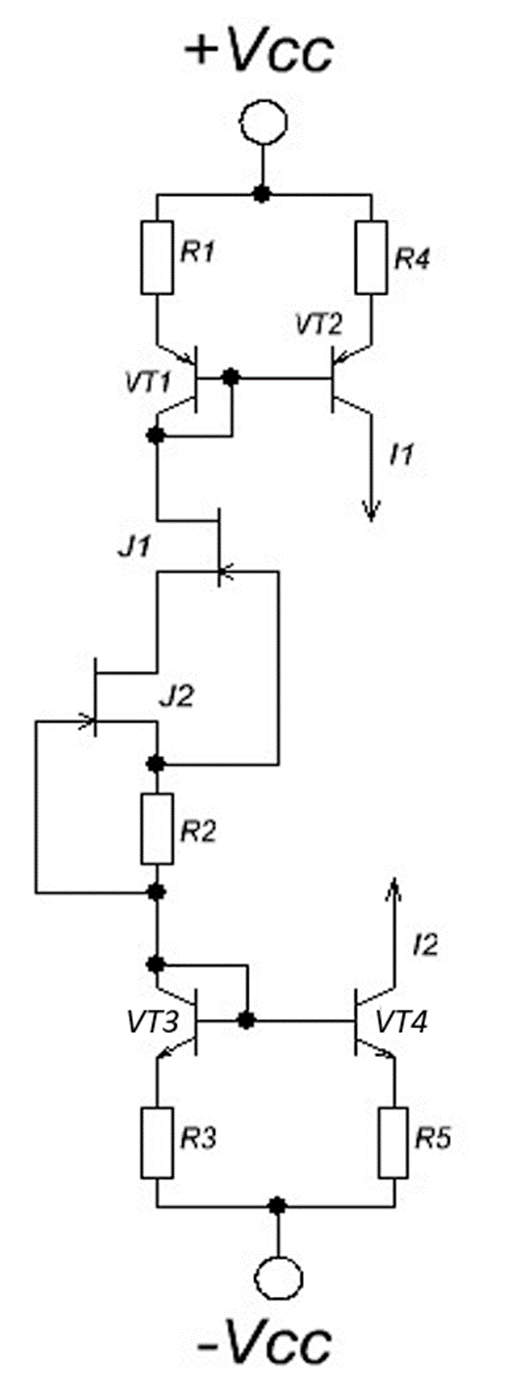}
\caption{Electrical scheme used for calculations. Vcc – supply voltage; R1-R5 – resistors; J1, J2 – FET transistors; VT2, VT3, VT4 – npn transistors; I1, I2 – currents.}
\label{Elscheme0}
\end{figure}

Third, our FET transistors have a large spread in gate cut-off voltages $V_{gs0}$, ranging from -0.5V to -2.5V. This leads to an inability to set the required reference current on crystals with extreme $V_{gs0}$ values, resulting in a low yield of good chips. To demonstrate this assertion, we calculate the value of resistor $R_2$ (Figure \ref{Elscheme0}) needed to obtain a 20 µA current flowing through the FET transistor J2: $I_d = I_{d0} (1 - V_{gs}/V_{gs0})^2$, where $I_d$ is the drain current, $I_{d0}$ is the drain current when the gate-source voltage is 0, $V_{gs}$ is the gate-source voltage, and $V_{gs0}$ is the gate cut-off voltage. $V_{gs} = - I_d R_2$; we get: $I_d = I_{d0} (1 + I_d R_2/V_{gs0})^2$. The $I_{d0}$ values obtained from the test FET transistors are as follows: for transistors with $V_{gs0}$ = -0.5 V, $I_{d0}$ is 32 $\mu$A. For transistors with $V_{gs0}$ = -2.5 V, $I_{d0}$ is 1600 $\mu$A. Substituting the values of $V_{gs0}$ and $I_{d0}$, we get $R_2$ values of 5.2 k$\Omega$ ($V_{gs0}$ = -0.5 V) and 110 k$\Omega$ ($V_{gs0}$ = -2.5 V). This resistance range exceeds the adjustment capabilities of resistors available in our technological line. 

Fourth, at drain-gate voltages ($V_{dg}$) greater than 22 V$\ldots$25 V, the drain-gate reverse current increases sharply. It can exceed 1 $\mu$A, which is unacceptable for input transistors. Therefore, an additional circuit to limit Vdg is needed for the input transistors. 

Fifth, it was found that produced thin-film resistors had poor characteristics, therefore they should be replaced by ion-implanted resistors \cite{Qian2005}.

Based on the conclusions outlined before, we proposed several modifications for the electric scheme of AD820 that would allow us to reach similar performance, but using strengths and limits of our production facilities. Changes were proposed for the input stage, second (pre-final) stage, output stage, and current reference.

\section {Proposed design of operational amplifier aRD820}
In Microsim software, we developed a model of the AD820 electrical schematic and configured the parameters of the electronic components within limits that can be reached by our production line, e.g. parameters for FET transistors. The simulation results showed that we could not achieve the desired parameters that correspond to the technical data of AD820 \cite{AD820_datasheet}. Therefore, several modifications to the electric scheme of AD820 were proposed and tested. Finally, we got an electric scheme for operational amplifier aRD820 with good simulation performance that corresponds to the targeted parameters (see Table \ref{Table1}). Below, we describe several modules of the aRD820 electrical schematic and explain why its performance surpasses that of the AD820 schematic. It should be emphasized that the AD820 electrical schematic is well-suited for the production capabilities of Analog Devices, which manufactures the AD820, that allows to reach good performance. However, due to the limitations of our production line, this electric scheme does not allow us to reach the expected performance. 

\subsection{Modification of Input stage module}
The input stage of an op-amp is designed to receive the input signal and provide initial amplification. This stage is typically implemented as a differential amplifier, which amplifies the voltage difference between the inverting and non-inverting inputs. One of the key functions of this stage is to reject common-mode signals, ensuring that only the differential signal is amplified. The input stage is crucial for maintaining a high input impedance, which minimizes the loading effect on the preceding circuit. In addition, the bias circuit within the input stage ensures that the transistors operate in the correct region, thus stabilizing the overall performance.

\begin{figure*}
\centering
\includegraphics[width=0.8\textwidth]{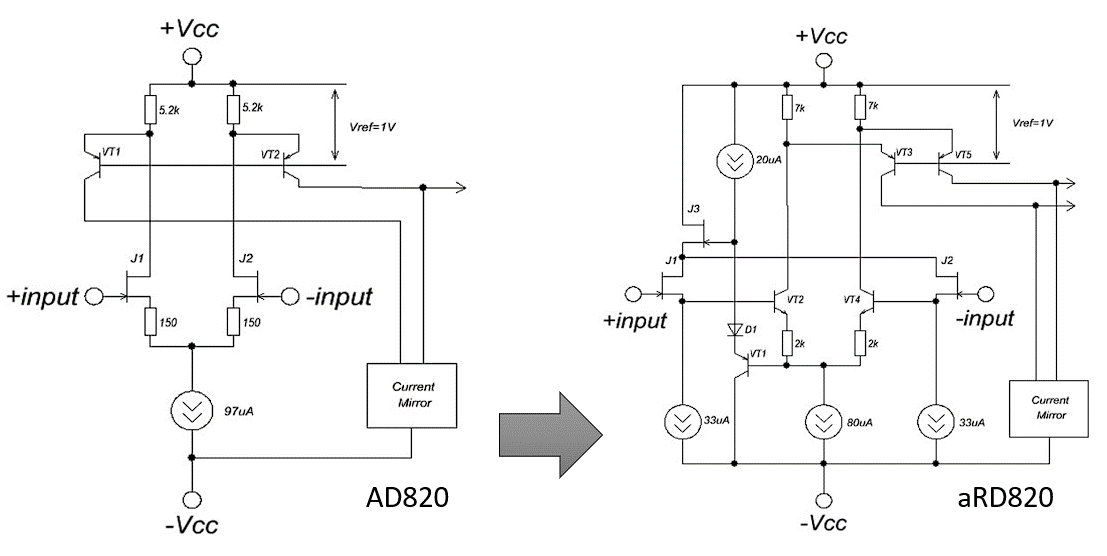}
\caption{A simplified electric schemes of an input stage modules of AD820 and aRD820}
\label{InputStage}
\end{figure*}

\begin{figure}
\centering
\includegraphics[width=0.5 \textwidth]{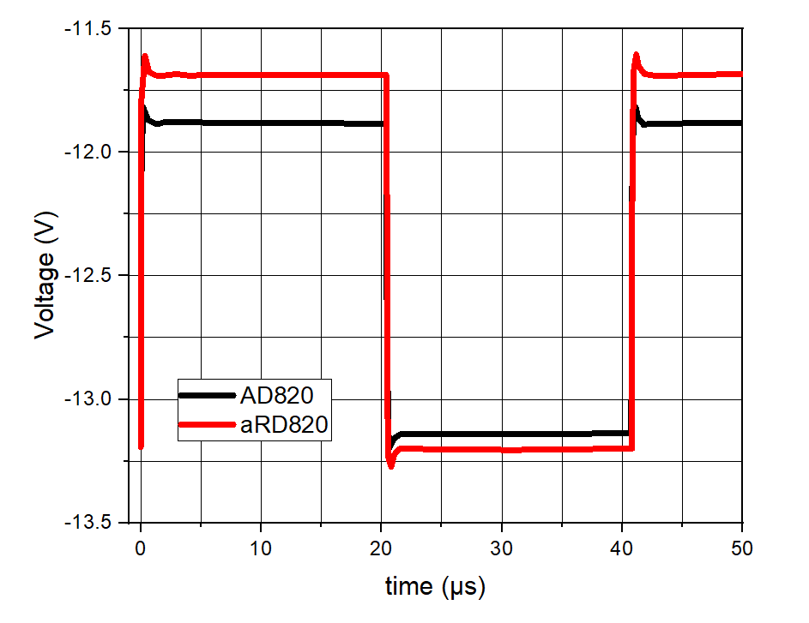}
\caption{Signal response of input stage modules of AD820 and aRD820. Vcc = $\pm$15 V, V(+inp) = 10 V, V(-inp) = 9.9 V $\ldots$ 10.1 V. Black line: Voltage on the collector of the  transistor VT2 of AD820 scheme (Figure \ref{InputStage}), Vout(p-p) = 1.25 V. Red line: Voltage on the collector of transistor VT5 of aRD820 scheme (Figure \ref{InputStage}), Vout(p-p) = 1.5 V.}
\label{Input10V}
\end{figure}

\begin{figure}
\centering
\includegraphics[width=0.5 \textwidth]{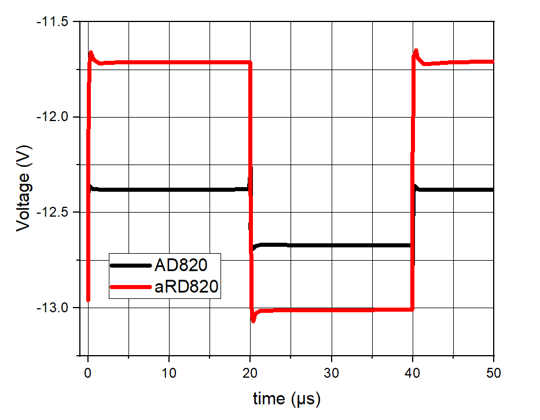}
\caption{Signal response of input stage modules of AD820 and aRD820. Vcc = $\pm$15 V, V(+inp) = 13 V, V(-inp) = 12.9 V $\ldots$ 13.1 V. Black line: Voltage on the collector of transistor VT2 of AD820 scheme (Figure \ref{InputStage}), Vout p-p = 0.33V. Red line: Voltage on the collector of transistor VT5 of aRD820 scheme (Figure \ref{InputStage}), Vout p-p = 1.3 V.}
\label{Input13V}
\end{figure}

Figure \ref{InputStage} shows simplified circuits of the AD820 and aRD820 input stages. By modifying the aRD820 input stage circuit relative to the AD820, we attempted to solve three problems. First, reduce the noise of the input stage. Second, expand the input signal range toward voltages close to +Vcc. Third, limit and stabilize the Vds voltage of the input transistors. As indicated above, implementing the input stage as given for AD820 could result in an input noise voltage swing (Vinp-p) of 5 $\mu$V to 10 $\mu$V in the range of 0.1 H to 10 Hz. Therefore, it was proposed to place source followers J1, J2 at the input of the operational amplifier. This configuration should reduce noise by five times or more. At the same time, using source followers and a differential pair of npn transistors VT2 and VT4  allows raising the upper limit of the allowable common-mode input signal by at least 0.4 V (since the voltages at the drains of J1 and J2 in these cases are +Vcc - Vref + Vbe for AD820 and +Vcc for aRD820).

Figures \ref{Input10V} and \ref{Input13V} illustrate the described advantage of the circuit for aRD820 in Figure \ref{InputStage}, obtained from computer models for input signals close to +Vcc. Instead of a current mirror, a resistive load of 50 k$\Omega$ was used, and the FET transistors' gate cut-off voltage was assumed to be Vgs0 = -2 V. In Figure \ref{Input10V} signal response of input stage modules of AD820 and aRD820 schemes are presented. Parameters used in simulations were Vcc = ±15 V, V(+inp) = 10 V, and V(-inp) is changed from 9.9 V to 10.1 V. Black line corresponds to the voltage Vc(VT2) on the collector of the transistor VT2 of AD820 scheme (Figure \ref{InputStage}) giving peak-to-peak value Vout(p-p) = 1.25 V. Red line corresponds to the voltage Vc(VT5) on the collector of the transistor VT5 of aRD820 (Figure \ref{InputStage}) giving peak-to-peak value Vout(p-p) = 1.5 V. In Figure \ref{Input13V} signal response of input stage modules of AD820 and aRD820 schemes are presented. Parameters used in simulations were Vcc = ±15 V, V(+inp) = 13 V, and V(-inp) is changed from 12.9 V to 13.1 V. Black line corresponds to the voltage Vc(VT2) on transistor VT2 of AD820 (Figure \ref{InputStage}) giving peak-to-peak value Vout(p-p) = 0.33 V. Red line corresponds to the voltage Vc(VT5) on transistor VT5 of aRD820 (Figure \ref{InputStage}) giving peak-to-peak value Vout(p-p) = 1.3 V. Figures \ref{Input10V} and \ref{Input13V} show that for the change of the input signal from +13 V to + 10 V, the gain of the circuit of AD820 drops four times, while the gain drop of the circuit of aRD820 is only 15$\%$. Similar simulations were made with parameters Vcc = ±15 V,  V(+inp) = 14.5 V and V(-inp) change from 14.4 V to 14.6 V. Then, voltage Vc(VT2) on the collector of the transistor VT2 of AD820 showed peak-to-peak value Vout(p-p) = 26 mV. Correspondingly, voltage Vc(VT5) on the collector of the transistor VT5 of aRD820 showed peak-to-peak value Vout(p-p) = 28 mV. 

At high supply voltages (+Vcc = +30 V; -Vcc = 0 V) and low input signals (Vin $<$ 7 V), there is a significant reverse current at the drain-gate junctions of the input transistors J1, J2 of (Figure \ref{InputStage}, electrical scheme AD820), causing the input current of the operational amplifier to exceed the norm. To avoid this effect, the aRD820 uses a circuit limiting the voltage in the drains of the input transistors, including the transistors VT1, J3, the diode D1 and a current source of 30 $\mu$ A. This ensures that the voltages at the drains of J1 and J2 are approximately equal to Vin - 2Vgs +Vbe, where Vgs is the gate-source voltage of J1 (J2). The average Vgs is -1.2 V and Vbe is 0.65 V. Therefore, the drain voltages of J1 and J2 will be Vin + 3.05 V, and Vdg at the input signals Vin $<$ +Vcc - 3.05 V remain almost constant at 3.05 V.

The input stage circuit of aRD820 has two drawbacks. First, the introduction of an additional stage increases the phase shift and theoretically reduces the stability. Second, the introduction of an additional stage imposes a restriction on the gate cut-off voltage Vgs0 of the input transistors J1 and J2. The gate cut-off voltage must not exceed -0.8 V $\ldots$ -0.9 V (i.e., it must be between -0.9 V and -2.5 V). Otherwise, the differential stage transistors VT2, VT3 will close when an input signal close to -Vcc is applied. Considering the advantages and disadvantages of the input stage circuit of aRD820, we deemed it feasible to use this circuit, addressing stability issues by modifying the pre-final stage, using additional frequency compensation in the final stage, and selecting wafers based on the Vgs0 parameter of test FET transistors during manufacturing.

During the design of the input stage, special attention was paid to minimizing the offset voltage of the operational amplifier. In AD820, it does not exceed 800 $\mu$V and is achieved by laser trimming of thin-film resistors located in the sources of the input FET transistors. Since we decided not to use thin-film resistors, we proposed to use an array of cutting resistors with fuses, blown by current, in aRD820. The trimming resistors are located in both arms of the lower npn transistors of the current mirror of aRD820. The resistor values are chosen so that the adjustment range of the operational amplifier offset voltage is from -10 mV to +10 mV, with the total voltage drop across the resistors not exceeding 100 mV.

\subsection{Modification of a second (pre-final) stage module}

Typically, the second stage, also known as the gain stage, provides most of the op-amp’s voltage gain. This stage may employ a common emitter or common source amplifier configuration, which enables high voltage amplification. The high gain of this stage allows the op-amp to amplify even very small input signals effectively. Additionally, this stage often includes frequency compensation, typically in the form of a Miller compensation capacitor, which helps to stabilize the amplifier by controlling the bandwidth and preventing oscillations. This ensures that the op-amp remains stable across a wide range of frequencies. For our scheme of aRD820, the main gain of op-amp was performed at the first stage. 

\begin{figure*}
\centering
\includegraphics[width=0.8\textwidth]{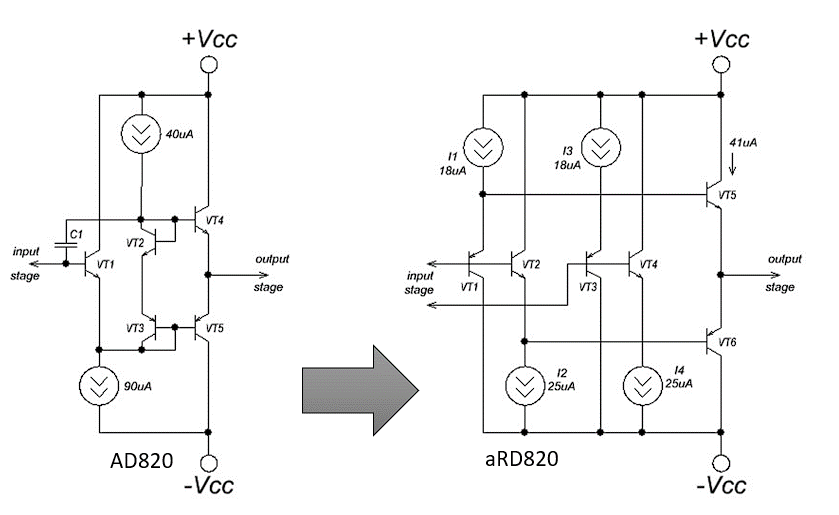}
\caption{A simplified electric schemes of the second (pre-final) stage modules of AD820 and aRD820}
\label{SecondStage}
\end{figure*}

Figure \ref{SecondStage} shows simplified circuits of the second (pre-final) stages of AD820 and aRD820. The second stage of aRD820 has a symmetrical structure, resulting in equal signal propagation times in the upper and lower arms. In AD820, the additional signal delay in the upper arm (VT1-VT3-VT2-VT4) is compensated by the parallel connection of capacitor C1. Calculations showed that the signal delay (and, consequently, phase shift) in the circuit in for aRD820 scheme is 3.5 times less than in the circuit in AD820 scheme, even at lower operating currents in the stage. Thus, the circuit of aRD820 will provide greater stability to the operational amplifier compared to the circuit of AD820. Additionally, the circuit of aRD820 uses additional emitter followers VT3, I3, and VT4, I4 to reduce the operational amplifier's offset voltage. They ensure identical input currents of the stage for any $\beta$ values of the npn and pnp transistors.

\subsection{Modification of output stage module}

The output stage of operational amplifier is responsible for driving the load connected to the op-amp's output. To achieve this, the output stage often uses a push-pull configuration, which can both source and sink current, making it highly efficient and capable of driving low-impedance loads. This stage provides current gain and reduces output impedance, ensuring that the op-amp can deliver sufficient current to the load without significant distortion. In some designs, a buffer is also used in the output stage to isolate the high-gain second stage from the load, further improving performance.

\begin{figure*}
\centering
\includegraphics[width=0.8\textwidth]{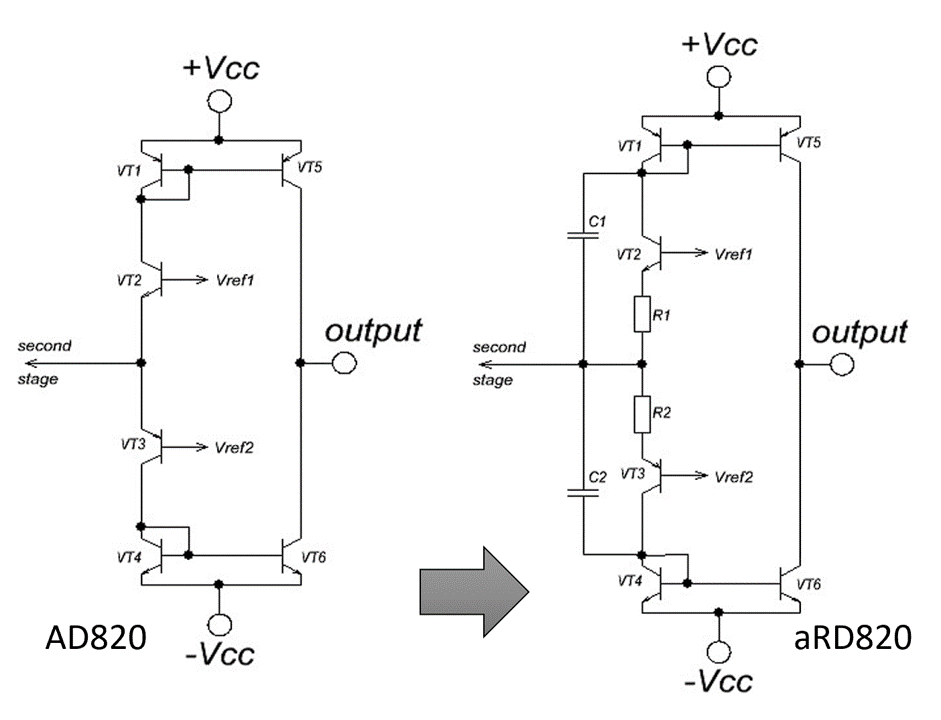}
\caption{A simplified electric schemes of the output stage modules of AD820 and aRD820}
\label{OutputStage}
\end{figure*}

Figures \ref{OutputStage} shows simplified circuits of the output stages of AD820  and aRD820. To compensate for the increased phase shift introduced by the VT2 and VT3 transistors in the aRD820 input stage (see Figure \ref{InputStage}), it was proposed to use frequency compensation in the output stage. The compensation elements are C1, C2, R1, and R2.

\begin{figure}
\centering
\includegraphics[width=0.5 \textwidth]{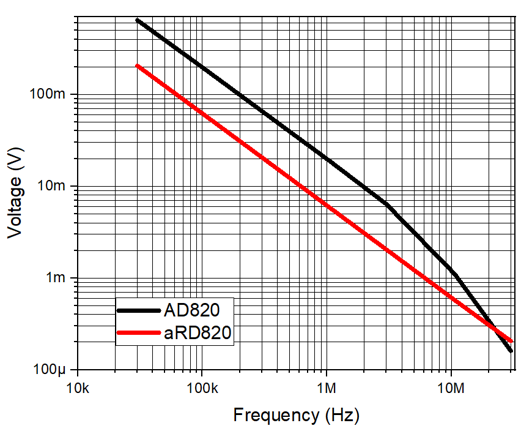}
\caption{Simulation of output module of operation amplifiers AD820 and aRD820. Voltage vs. frequency}
\label{OutputStage-VoltageFreq}
\end{figure}

\begin{figure}
\centering
\includegraphics[width=0.5 \textwidth]{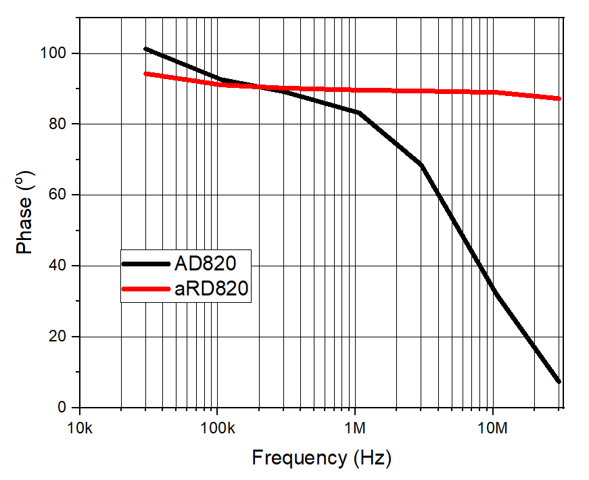}
\caption{Simulation of output module of operation amplifiers AD820 and aRD820.Phase vs. frequency}
\label{OutputStage-PhaseFreq}
\end{figure}

Figures \ref{OutputStage-VoltageFreq} and \ref{OutputStage-PhaseFreq}  illustrate a comparative analysis of the amplitude-frequency and phase-frequency characteristics of simulated signals of the circuits AD820 and aRD820. The parameters used in simulations were Vcc = $\pm$15V, VACinp = 1mV, Rload = $\infty$, and Cload = 0. The gain of the output stage of aRD820 is lower than that of the circuit in AD820, but we considered this acceptable due to the ample margin for this parameter. The phase-frequency characteristic of the circuit of aRD820 appears preferable to that of AD820 scheme. The phase margin of the circuit of AD820 at 3MHz is 68.6 degrees, while for aRD820 it is 89.6 degrees. The frequency of 3 MHz was chosen because it is close to the unity gain frequency of the operational amplifier. The operational amplifier is considered stable with a phase margin of at least 45 degrees at the unity gain frequency. 

\subsection{Modification of current reference module}
The current reference circuit is a fundamental component that ensures stable operation of the op-amp under varying conditions, such as changes in temperature or power supply voltage. The current reference typically consists of a constant current source and current mirrors, which provide a stable bias current to the transistors in both the input and second stages. This stable biasing ensures consistent operation, allowing the op-amp to maintain its performance across a range of operating conditions.

\begin{figure*}
\centering
\includegraphics[width=0.8\textwidth]{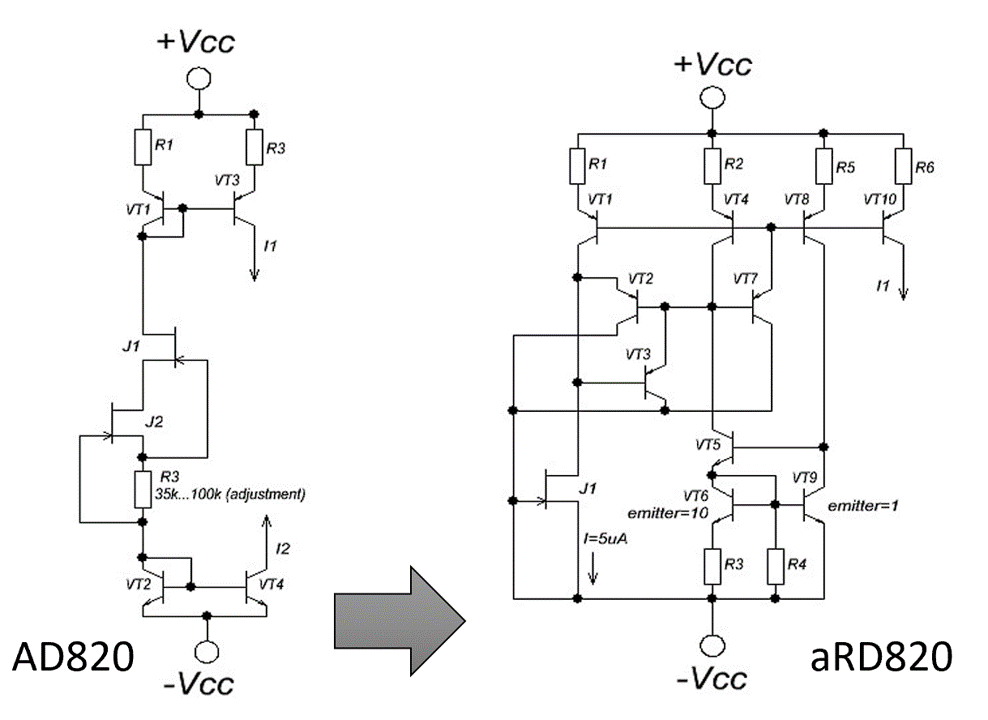}
\caption{A simplified electric schemes of the current reference modules of AD820 and aRD820}
\label{CurrentReference}
\end{figure*}
As described earlier, using a reference source circuit of AD820 would inevitably result in a low yield of good op-amp chips. Therefore, it was proposed to eliminate the FET transistors determining the reference current magnitude. Figure \ref{CurrentReference} shows the reference source circuits of the AD820 and aRD820. 

In the current reference circuit of aRD820, there is only one FET transistor that is used as an element to initiate the circuit, so it can also work with average characteristics. The reference current is set using the pnp transistor VT10. This current is approximately equal to the sum of the currents through resistors R3 and R4. The resistance of the diffusion (or ion-implanted) resistors depends on temperature. The current through resistor R3 has a positive temperature coefficient, while the current through resistor R4 has a negative one.  Thus, the resistances of R3 and R4 are chosen so that the current through VT10 remains constant over a temperature range of -40 $^o$C to +85 $^o$C. The exact reference current value will be set using trimming resistors connected in series with R6. The total drift of the reference current over the temperature range of -40 $^o$C to +85 $^o$C and the entire supply voltage range should not exceed 5$\%$ to 10$\%$.

\subsection{Simulations of the full electric schemes of AD820 and aRD820}
\begin{figure}
\centering
\includegraphics[width=0.5 \textwidth]{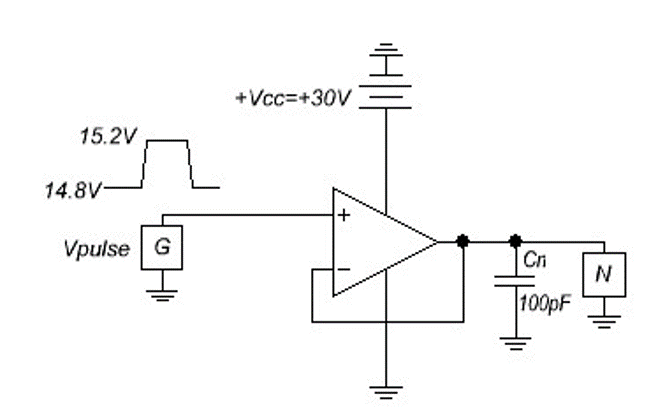}
\caption{Schematics to test the performance of operation amplifiers AD820 and aRD820}
\label{TestScheme}
\end{figure}

The modified components described above and the complete operational amplifier circuit were simulated using the Microsim circuit simulation program. Parameters of the components and the entire operational amplifier model were determined over temperature ranges (-40 $^o$C to +85 $^o$C), supply voltages (+30 V 0V; +5 V 0 V; $\pm$2.5 V; $\pm$ 5V; $\pm$ 15V), input signals (-Vcc $\ldots$ +Vcc), load capacitances (up to 350 pF), and output currents (up to 20 mA). Particular attention was paid to the impact of replacing thin-film resistors (as in AD820) with diffusion and ion-implanted resistors, as diffusion and ion-implanted resistors have significantly higher temperature coefficient of resistance. Simulation showed that this replacement did not significantly affect the operational amplifier parameters. Moreover, special attention was given to the model's dynamic properties, specifically its stability. To assess stability, an operational amplifier configuration with a gain of A = +1 and a load capacitance of Cn = 100 pF was used (Figure \ref{TestScheme}).

\begin{figure}
\centering
\includegraphics[width=0.5 \textwidth]{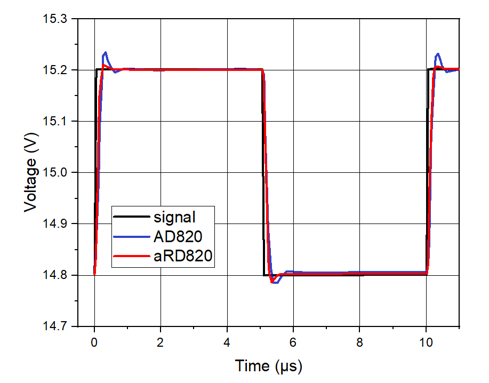}
\caption{Input and output signal oscillograms (simulations)}
\label{OutputSignal_simulation}
\end{figure}

\begin{figure}
\centering
\includegraphics[width=0.5 \textwidth]{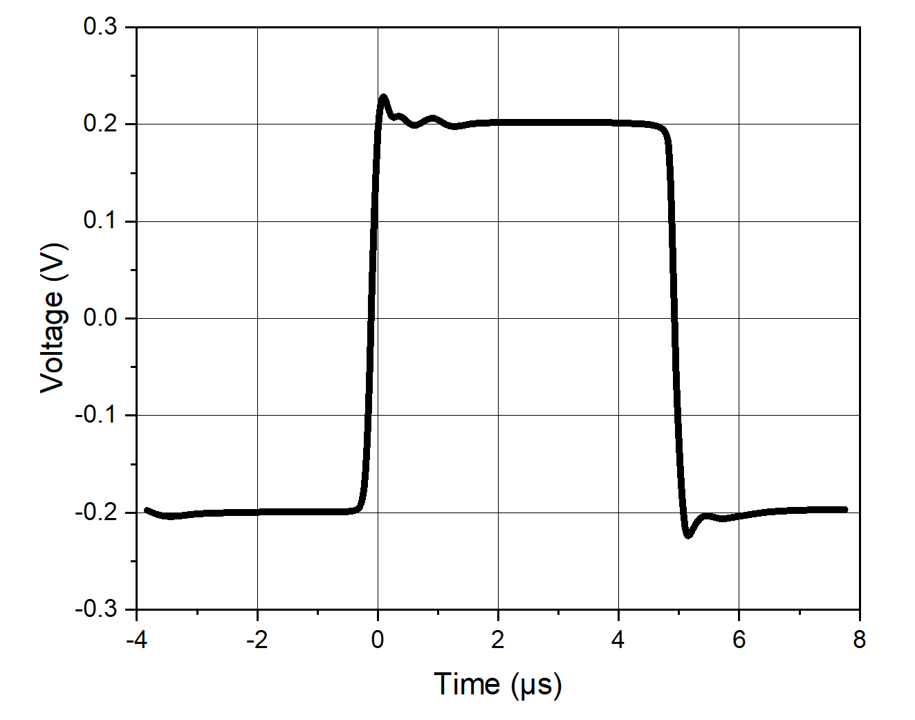}
\caption{Output signal oscillograms (tests)}
\label{OutputSignal_test}
\end{figure}

Figure \ref{OutputSignal_simulation} shows the input and simulated output signal oscillograms of the AD820 and aRD820 model for the above configuration.  As seen, the AD820 has a longer transition process during voltage jumps. Additionally, the aRD820 has a slightly higher slew rate. Finally, Figure \ref{OutputSignal_test} shows the output signal oscillogram of the actual aRD820 microchip (Vcc = $\pm$15 V) that demonstrates similar performance. 

Later production and tests of aRD820 chips showed that the expected parameters of the operational amplifier given in Table \ref{Table1} are achieved.  

\section{Conclusion}
The task of the research was to design, construct and test a single channell low noise rail-to-rail operation amplifier aRD820 with the specification given in Table 1, based on an initial prototype of Analog Device AD820 chip. Due to limitations and specificity of available production lines, the construction of a chip that has implemented an electric scheme of AD820 prototype is not reasonable as a high rate of damaged chips with poor performance will be obtained. Therefore, a modified electric scheme of AD820 prototype was proposed that allows it to reach the targeted performance. The input stage module got source followers at the input of the operational amplifier (Figure \ref{InputStage}). 
Second stage module was modified to be more symmetric (Figure \ref{SecondStage}). The output stage module obtained additional resistors and capacitors to achieve a frequency compensation (Figure \ref{OutputStage}). One FET transistor in the current reference module was substituted by other elements (Figure \ref{CurrentReference}). The performance of modified electric schemes of modules was tested in Simulink software. Simulations of the full electric scheme for aRD820 were made and showed that it demonstrates similar characteristics as AD820 data tables. The aRD820 chip was produced, and measurements demonstrated that the planned characteristics of the operational amplifier were met.

\section*{Acknowledgements}
Research activities were funded by Latvian Recovery and Resilience Mechanism Plan under reform and investment direction 5.1.r. "Increasing productivity through increasing the amount of investment in R$\&$D", reform 5.1.1.r. “Management of innovations and motivation of private R$\&$D investments", investment 5.1.1.2.i. "Support instrument for the development of innovation clusters",  project No. 5.1.1.2.i.0/1/22/A/CFLA/002 "Competence Center of the Latvian Electrical and Optical Equipment Manufacturing Industry", within specific research project No. 1.6. “Research and Development of Low Voltage and Low Noise Single-Stage Rail-to-Rail Operational Amplifier Technology and Chip Prototype”.

\section*{Author contributions}
Conceptualization, S.R., M.L., and D.K.; methodology, S.R., M.L. and D.K.; software, D.K.; validation, S.R.; formal analysis, D.K., S. R., M.L. and A.A.; investigation, D.K. and S. R.; data curation, D.K.,  S.R. and A.A.; writing - original draft preparation, D.K., A.A.; writing - review and editing, D.K., A.A., S. R. and M. L.; visualization, D.K. and A.A.; project administration, M.L.; funding acquisition, M.L. All authors have read and agreed to the published version of the manuscript.

%\subsection{Example Subsection}
%\label{subsec1}

%Subsection text.

%% Use \subsubsection, \paragraph, \subparagraph commands to 
%% start 3rd, 4th and 5th level sections.
%% Refer following link for more details.
%% https://en.wikibooks.org/wiki/LaTeX/Document_Structure#Sectioning_commands

%\subsubsection{Mathematics}
%% Inline mathematics is tagged between $ symbols.

%\begin{eqnarray}
% f(x) &=& (x+a)(x+b) \nonumber\\ %% If equation numbering is not %needed for a row use \nonumber.
%      &=& x^2 + (a+b)x + ab
%\end{eqnarray}

%% Refer following link for more details.
%% https://en.wikibooks.org/wiki/LaTeX/Mathematics
%% https://en.wikibooks.org/wiki/LaTeX/Advanced_Mathematics

%\appendix
%\section{Example Appendix Section}
%\label{app1}


%%       \cite{<label>} ==> [1]

%%
%Example citation, See \cite{lamport94}.



% Use elsarticle-num.bst as the bibliography style
\bibliographystyle{elsarticle-num}
\bibliography{biblio}  % Your .bib file, without %\begin{thebibliography}{00}

%% For numbered reference style

%\bibitem{lamport94}
%  \textit{\LaTeX: a document preparation system},
%  Addison Wesley, Massachusetts,

%\end{thebibliography}
\end{document}